\documentclass[amsmath,amssymb,12pt,floatfix] {revtex4}
\usepackage[dvips]{graphicx}
\usepackage{bm}
\begin{document}
\title{Importance of boundary effects in diffusion of hydrocarbon 
molecules in a one-dimensional zeolite channel}
\author{Sakuntala Chatterjee and Gunter M. Sch\"{u}tz}
\affiliation{Institut f\"{u}r Festk\"{o}rperforschung,
Forschungzentrum J\"{u}lich, D-52425 J\"{u}lich, Germany}
\begin{abstract}
{\it Abstract:} Single-file diffusion of propane and toluene molecules
 inside a narrow, effectively one-dimensional zeolite pore was experimentally
studied by Czaplewski {\sl et al.} Using a stochastic lattice gas approach,
 we obtain
an analytical description of this process for the case of single-component
 loading. We show that
a good quantitative agreement with the experimental data for the desorption
temperature of the hydrocarbon molecules
can be obtained if the desorption process from the boundary is associated with a
higher activation energy than the diffusion process in the bulk. We
also present Dynamical Monte Carlo simulation results for two-component
loading which demonstrate in agreement with the experimental
findings the effects of mutual blockage of the
molecules due to single-file diffusion. 
\end{abstract}
\maketitle
\section{Introduction}
Zeolites are microporous crystalline materials which have wide industrial 
applications. Because of their regular pore structure of molecular dimension
zeolites are often used as `molecular sieves' to selectively sort molecules 
based on their size (or shape).  
In a wide variety of chemical and petro-chemical processes, 
zeolite channels are used as catalysts and adsorbents of hydrocarbon
molecules. It is important to understand the mechanism of transport of
molecules within  a zeolite channel and their exchange with the surrounding
gas-phase in order to design more efficient use
for such materials.

In this paper we describe a lattice gas model to explain the mechanism
of transport and desorption of hydrocarbon molecules in a 
quasi one-dimensional zeolite channel. In particular,
we aim to explain quantitatively the experimental observation of single-file
diffusion by Czaplewski {\sl et al.}~\cite{snurr}. An earlier study shows
that, using activated diffusion of hard-core particles on a one-dimensional
lattice, one can explain the important features of the
experimental data qualitatively~\cite{sg}. However,
 no quantitative comparison was
possible within that simple model. In this paper we address the question:
Which physical mechanism is responsible for this failure? We find that 
 boundary effects
in the form of a higher activation energy for desorption of molecules play a
crucial role. With judiciously chosen desorption barriers we can match our
analytical and numerical results quantitatively with the experimental data. 
In the remaining part of the introduction, we present the  
main idea of the experiment in~\cite{snurr} and we briefly illustrate our 
modeling strategy.

\begin{figure}
\includegraphics[scale=0.7,angle=0]{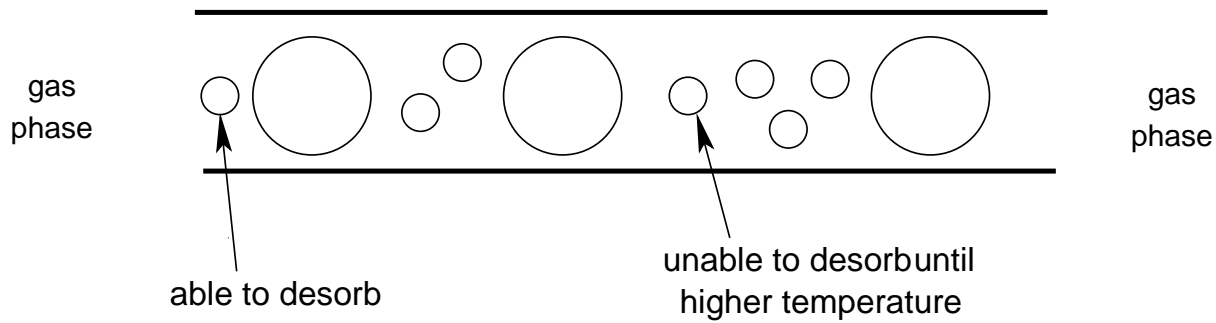}
\caption{ Schematic diagram of $1$-d zeolite pore to show the trapping of
light, less-strongly adsorbed molecules by heavier, more-strongly adsorbed
ones. }
\label{fig:pore}
\end{figure}

 In~\cite{snurr} Czaplewski {\sl et al.}
have demonstrated in an experiment that it is possible to trap the
light hydrocarbon (propane) 
molecules in presence of the heavier ones (toluene) inside a
narrow zeolite channel. Toluene molecules are strongly adsorbed in
the zeolite and consequently needs a high temperature in order to desorb from
the channel, whereas the propane molecules,
 being weakly adsorbed, have a rather low
desorption temperature. If a mixture of toluene and propane 
is present inside a narrow channel of the zeolite of type EUO, then, because of the confining pore
dimension, the molecules cannot pass each other (see Fig.\ref{fig:pore}). In such an
effectively one-dimensional motion of the molecules the
more strongly adsorbed toluene molecules block the movement of the less
strongly adsorbed propane molecules.
 As a result, propane is not able
to desorb until toluene has desorbed, which occurs at a high
temperature. Thus the effective desorption temperature of propane
is raised in the presence of toluene.

We aim to describe the main features of the above experimental observation
using a stochastic lattice gas approach. Following
 K\"arger {\sl et al.}\cite{jk1,jk2}, we model
the narrow pore of the zeolite using a one-dimensional lattice and the
hydrocarbon molecules as a set of hard-core particles
diffusing on that lattice. As shown in~\cite{sg}, this simple picture
can indeed explain the main experimental results qualitatively. Using an
Arrhenius form for activated diffusion of the 
hard-core particles on the lattice, we have been able to show that the
presence of strongly adsorbed particles raises the desorption temperature of
the weakly adsorbed ones, in conformity with the experimental observation
in~\cite{snurr}. However, this approach
 predicts a desorption temperature which is
much higher than  the experimentally observed value.

In order to achieve progress, we suggest that 
the evaporation of molecules at the boundary of the lattice 
is associated with a larger activation energy than the diffusion at the bulk.
In a real system, the
presence of such a desorption  barrier near the exit of the channel to the gas phase
can be
explained by considering the attractive van der Waals interaction between the
molecules and the pore wall. When a molecule leaves the pore of a zeolite, it
loses its close proximity with the lattice atoms and moves out into a
low-pressure gas phase, thereby giving up favorable dispersion
energy~\cite{arya,vasenkov}.

In the next section we give a brief description of the
experiment~\cite{snurr} and of our previous work on modeling the experimental
setting~\cite{sg,sg2}. In section $3$ we consider the presence of the
desorption barrier. We present an analytical approach to study
single-component loading to demonstrate how the single-component desorption
profile changes after incorporating the desorption barrier.  
In section $4$ we present our
dynamical Monte Carlo simulation results for the two-component system.

\section{One-dimensional zeolites as hydrocarbon traps: Experiment and
modeling}

For the sake of self-containedness of this paper, we describe in this section
the experiment carried out by Czaplewski {\sl et al.} to demonstrate the use 
of narrow zeolite channels as hydrocarbon traps. In the second half of this
 section we summarize our earlier results obtained from the lattice gas 
approach that we develop further in the bulk of the paper.

\subsection{Outline of the experiment by Czaplewski {\sl et al.} }  
Several different zeolite samples with one-dimensional channel or with
three-dimensional pore-network connectivity were considered. The zeolite
samples were  loaded with an equimolar binary mixture of propane and
toluene and, for reference purposes, 
also with  single-component propane and toluene separately. After the loading
was complete, the whole system was purged in pure helium such that no
hydrocarbon molecules remain outside the channels. Then the system was heated
at a constant rate and the flow of the 
hydrocarbon molecules out of the channel was
 monitored as a function of temperature. This is known as `temperature
programed desorption' (TPD).

\begin{figure}
\includegraphics[scale=1.0,angle=0]{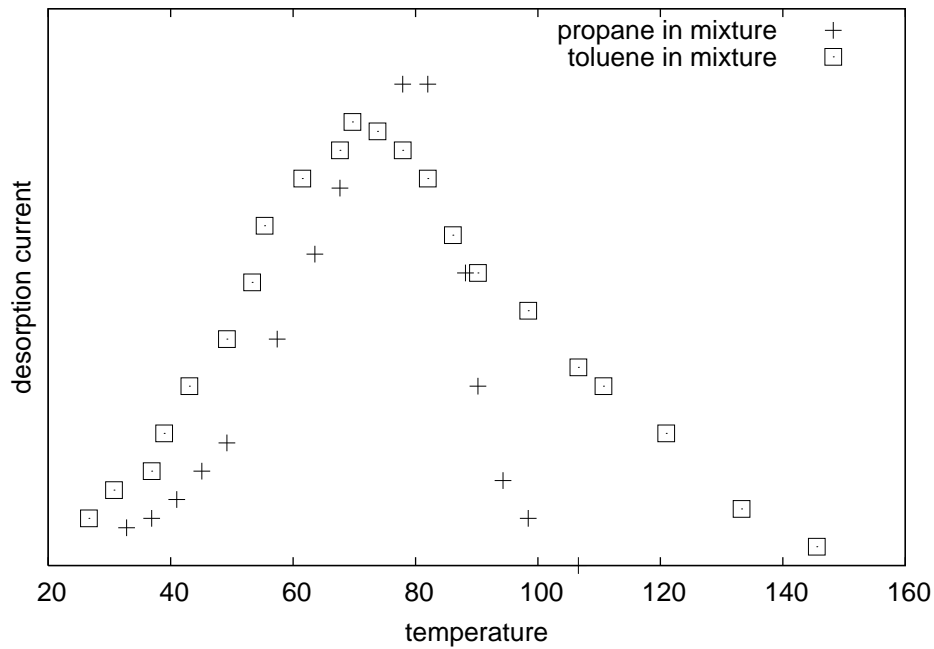}
\caption{Experimental data for the desorption profile
as a function of temperature for propane and toluene in a binary
mixture in the zeolite EUO, as measured by Czaplewski {\sl et al.} Data points
taken from Fig. $5$ of~\cite{snurr}.  }
\label{fig:fig5}
\end{figure}

For the single-component
loading, it was found that the current initially grows with temperature, shows
a peak and then falls off to zero. The desorption temperature of each
component was measured at the position of the peak. For the purpose of our
modeling we will be interested in the TPD profile of the EUO zeolite
(see~\cite{sg} for details). When only propane was loaded into
 the one-dimensional channel of EUO, the desorption peak was found
at the temperature $40^{\circ}$C. For single-component toluene the
desorption temperature was $80^{\circ}$C, toluene being more strongly
adsorbed. For an equimolar binary mixture of the two gases in EUO,
 the propane desorption peak is found to 
 occur at a substantially higher temperature ($75 ^{\circ}$C) and the toluene
desorbs at  $70 ^{\circ}$C, as shown in Fig. \ref{fig:fig5}. 

Thus, this experiment demonstrates that using zeolite samples like EUO with
one-dimensional channel it is possible to trap the light hydrocarbon
molecules until higher temperature. No such effect has been observed for
Na-ZSM-5 zeolite which has a three-dimensional pore-network connectivity. This
shows that the presence of single-file diffusion is primarily responsible for
the trapping effect observed in the experiment. 

\subsection{Lattice gas modeling}  

\begin{figure}
\includegraphics[scale=0.9,angle=0]{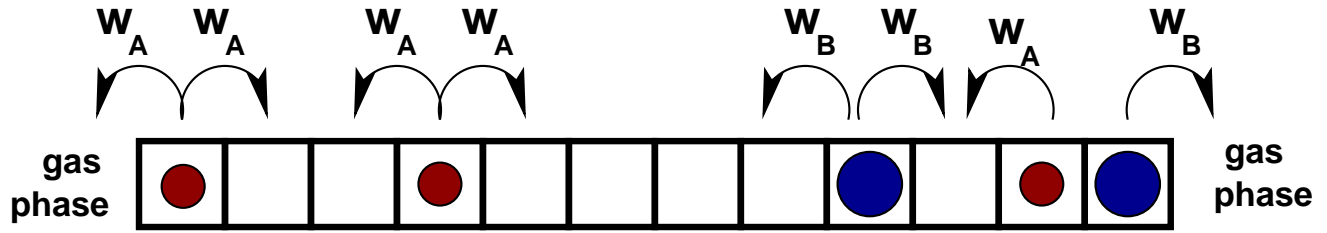}
\caption{ Two component SEP on an open lattice without boundary injection}
\label{fig:lattice}
\end{figure}

We model the narrow pore of
EUO by a one-dimensional lattice whose ends are open. 
 The diffusion of propane and toluene in the pore is modeled by a
two-component symmetric exclusion process (SEP) where hard-core particles of
two different species move on a lattice of $L$ sites~\cite{chou}. Propane
and toluene molecules are represented
as $A$ and $B$ particles, respectively. An $A$($B$) particle
can hop to the nearest neighbor site in either direction with rate
$w_A$($w_B$) if
the destination site is empty. Thus the particles have a hard-core exclusion
among themselves. At a boundary site of the lattice the $A$($B$)
particles can exit the system with rate $w_A$($w_B$).
Note that in this model we
do not consider any additional energy barrier for desorption. Both bulk
diffusion and boundary desorption occur with the same rate.
In Fig. \ref{fig:lattice} we have shown the possible dynamical moves
of the model~\cite{sg}. 

In this model we do not allow for boundary injection, {\sl i.e.},
once a particle
hops out of the lattice from a boundary site, it is immediately removed from
the system; no particle is allowed to enter the lattice through the boundary.
Such a boundary condition resembles the experimental scenario where after the
initial loading no more molecules are loaded into the zeolite sample in
the course of the TPD measurement.

To model the temperature programed
desorption carried out in  experiment, we increase the temperature $T$
 in our model uniformly with time with a heating rate $\lambda$ such that
\begin{equation}
 T(t)=T_0+\lambda t. 
\label{eq:lambda}
\end{equation}
The hopping rates are assumed to have an Arrhenius dependence on temperature:
\begin{eqnarray}
\nonumber
w_A = \Gamma _ A \exp \left ( -E_A/kT \right ) \\
\label{eq:arr}
w_B = \Gamma _ B \exp \left ( -E_B/kT \right ) 
\end{eqnarray} 
where $k$ is the Boltzmann constant and $T$ is the time-dependent temperature
in (\ref{eq:lambda}). In our model the $B$ particles are assumed
to be heavier and more strongly adsorbed. This means the activation energy for
diffusion is larger for $B$ particles, {\sl i.e.}, $E_B > E_A$. As shown below,
we set the overall time scale such that $w_A$ and $w_B$ are always less than unity.
 As a result, these rates can be directly interpreted as probabilities
 for Monte Carlo simulations. Notice that for single particles the hopping rates
multiplied with the square of the lattice constant are equal to the diffusion 
constants of these particles. Therefore we refer to these rates also as 
diffusivities.

In the two-component SEP with time-dependent hopping rates described above
 it is difficult to carry
out any analytical calculation. In the case of time-independent rates, it
is possible to describe the time-evolution of the density profile using a set
of coupled non-linear differential equations~\cite{brzank2,brzank}.
However, for time-dependent rates, this Maxwell-Stefan type
approach can be used only if the system is in local equilibrium which
we find hard to justify for the experimental
scenario of Ref~\cite{snurr}. So we
studied the system in~\cite{sg} using dynamical Monte Carlo simulation.
However, in the case of single-component loading, {\sl i.e.}, when only one
species of particles is present on the lattice, it is possible to
solve the system exactly and
obtain a closed form expression for the desorption profile~\cite{sg2}.

Let $\rho_x(t)$ denote the average occupancy at site $x$ at time $t$.
The time-evolution of
$\rho_x(t)$ is governed by the diffusion equation on a lattice~\cite{domb}:
\begin{equation}
\frac{\partial \rho_x(t)}{\partial t} = w_{\alpha}(t)\left ( \rho_{x+1}(t) +
\rho_{x-1}(t) - 2\rho_{x}(t) \right )
\end{equation}   
where $w_{\alpha}(t)$ is the time-dependent diffusivity as defined in  
(\ref{eq:arr}) with $\alpha$ being either $A$ or $B$, depending on which species
is present. In order to solve this lattice diffusion equation,
we reparametrize time as $d\tau = w_{\alpha}(t) dt$. In terms of this
reparameterized time $\tau$ the above equation becomes an ordinary diffusion
equation without any explicit time-dependence in the diffusivity.
This can be solved by using the ansatz 
\begin{equation}
\rho_x (\tau)= \sum_k \left ( A_k(\tau) e^{ikx}+B_k(\tau) e^ {-ikx} \right )
\end{equation}
where $A_k$ and $B_k$ are constants that depend on the boundary conditions and 
the initial density profile. 

Since there is no boundary injection
into the lattice, we use the boundary condition $\rho_x(\tau)=0$ for $x=0,L$.
The solution then turns out to be
\begin{equation}
\rho_x(\tau)= 2 \sum_{n=1}^{(L-1)/2} A_n \exp \left [ -2\tau \left ( 1- \cos 
\frac{(2n+1) \pi}{L} \right ) \right ] \sin  \frac{(2n+1) \pi x}{L} 
\label{eq:rho}
\end{equation} 
where the value of $\tau$ can be 
obtained by numerically performing the integration 
$\int_{0} ^ \tau dt w_{\alpha}(t)$. The choice of initial condition determines
the constant $A_n$. We consider a homogeneous initial condition 
$\rho_x(0) = \overline{\rho}$, for the bulk sites $x \neq 0,L$.
 This is based on the assumption that  
in the experiment, when the zeolite samples are loaded with
hydrocarbon molecules, the loading procedure gives rise to uniform equilibrium
bulk density. Such an initial condition yields
\begin{equation}
A_n =\frac{1}{2}\overline{\rho}L \sum_{x=1}^{L} \sin \left( \frac{(2n+1) \pi
x}{L}
\right) . 
\label{eq:an}
\end{equation}
The instantaneous desorption current $J_{\alpha} (t)$ is then given by
  $ w_{\alpha}(t)\left(
\rho_1(t)+\rho_{L-1}(t) \right)$, which can easily be evaluated using the
relations \ref{eq:rho} and \ref{eq:an}~\cite{sg2}.

\begin{figure}
\includegraphics[angle=0,scale=1.0]{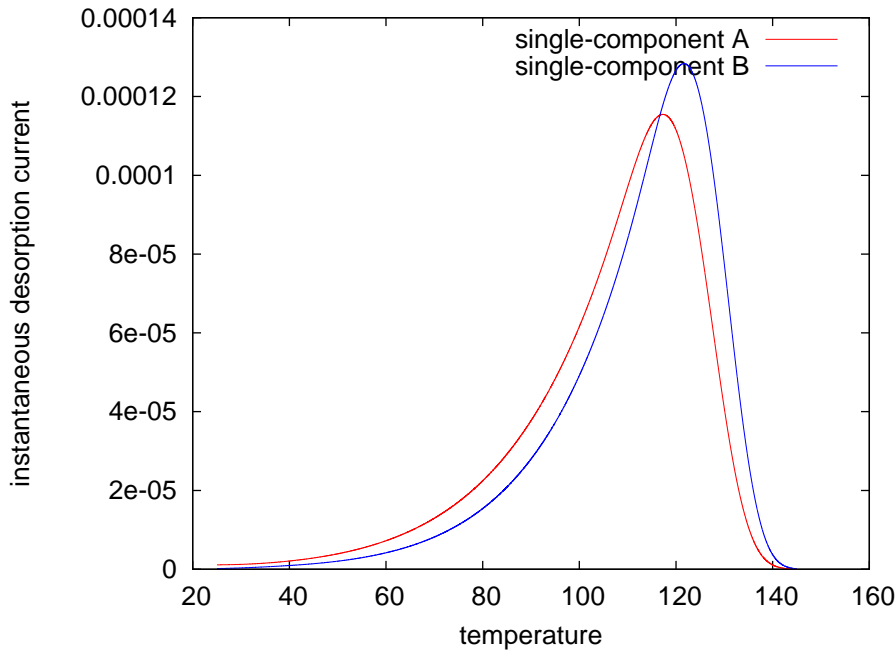}
\caption{Variation of the instantaneous desorption current computed using Eq.
\ref{eq:rho} and \ref{eq:an} as a function of temperature
 (in $^{\circ}$C) for single-component loading. }
\label{fig:single}
\end{figure}

Before we present our results, a brief discussion about the choice of
parameters is in order. 
The typical channel length of an EUO zeolite is a few $\mu m$ 
which is about few thousand times the size of the diffusing hydrocarbon
molecules. For computational simplicity, we work with a lattice size $L=1000$.
Also, in our calculations, we use 
the same temperature range $27-150 ^ \circ C$ as considered in the experiment
and have chosen the activation energy $E_{\alpha}$ and the heating
rate $\lambda$ such as to obtain a desorption peak within this
temperature range. We have used an increment rate $\lambda =10^{-5}$ degree
per unit time  to ensure that the current drops to zero at the final
temperature, as in the experimental setting.
When only $A$ particles are diffusing on the lattice, we use
$E_A=116.3$ kJ/mol. The factor  $\Gamma_A$, which sets the time-scale,
has been given a large value such that the variation of $w_A$
in the above temperature range is substantial which, as will be shown below, 
is necessary to reproduce the experimental data. This is
ensured by setting $\Gamma_A = \exp \left(E_A/kT_f \right) $, where $T_f$ is
the final temperature.  We present our result for $J_A(t)$ in Fig.
\ref{fig:single}. In the same figure, the desorption profile
 for the single-component loading of
$B$ particles is also shown, with $E_B=133$ kJ/mol,
and $\Gamma_B=\exp \left (E_B/kT_f \right )$. Note that the desorption peak
for $J_B(t)$ occurs at a higher temperature than
that for  $J_A(t)$. In other words, $B$ is more strongly adsorbed than $A$, as
expected. Note that $E_{\alpha}$ in this case 
is not a truly physical
quantity and its value depends on the details of the modeling approach.

Our calculation yields a desorption profile whose shape is
similar to that
seen in experiment. As temperature increases, the diffusivity grows and as a
result  $J_{\alpha} (t)$ also rises. On the other hand, 
 since there is no boundary
injection, the lattice starts getting depleted of particles and
after attaining a peak $J_{\alpha} (t)$ falls off.  
However, as shown in Fig. \ref{fig:single},
 the desorption peaks in our model occur at temperatures far too
high: In the experiment single-component propane and toluene peaks for EUO
occur at $40^{\circ}$ and $80^{\circ}$C, respectively~\cite{snurr}.
It is possible to change the 
peak position in our model by changing  $\lambda$ and
$E_{\alpha}$. A smaller $\lambda$ and/or a smaller
$E_{\alpha}$ would shift the peak towards lower temperature values. However, this
 would also
change the qualitative features of the profile. One can show that if
the values of $\lambda$, $E_{\alpha}$ are reduced $J_{\alpha} (t)$
starts from a large value and undergoes an
initial drop before it peaks again at the desorption temperature. Such an
effect is not observed in the experimental desorption profile. Thus it is not
possible within this model to obtain a quantitative agreement with the
experimental desorption temperature.

\section{Desorption Barrier and the Peak Position }

 Let us now consider which physical mechanism is missing to allow for
a realistic desorption temperature. This mechanism must be such that
the above mentioned initial
fall in the current is suppressed, while the peak can be moved to the left. We
suggest that this can be achieved by choosing a higher activation energy for the
exit rate at the boundary. In other words, the particles undergo an activated
diffusion in the bulk of the lattice with an activation energy $E_{\alpha}$,
as before, but the desorption from the boundary is associated with a higher
activation energy $V_{\alpha}$.

The presence of a higher activation energy near
the exit of a zeolite pore has been reported in earlier
studies involving molecular dynamics simulations~\cite{arya,vasenkov,bdry}.
The importance of such a surface energy barrier on the diffusion of adsorbed
molecules inside a zeolite channel was studied in~\cite{arya}. The diffusion
of methane molecules inside the narrow pore of $AlP0_4$-$5$ zeolite was
studied in the presence of a desorption barrier at the pore exit, 
using molecular dynamics simulations. It was found
that the effect of this pore-exit barrier becomes less and less pronounced as
the loading ({\sl i.e.}, the initial density of methane molecules inside the
pore) is increased. This was explained by considering the attractive interaction
between the molecules which gives rise to local clustering. Such clusters are
often stable and the energy barrier for cluster diffusion is often lower than
that of  single-molecule diffusion~\cite{sholl}. It was argued in~\cite{arya}
that near the pore exit the desorption of a molecule into the gas phase
is aided by the neighbors behind it,
pushing it over the energy barrier, thus effectively reducing the
activation barrier for desorption. However, at low loading
the molecules are too far apart and the escape of one
sorbate molecule is unaffected by the presence of other molecules. Therefore,
a molecule can exit the pore only when it has gathered enough momentum to jump
over the energy barrier at the pore exit.

In our modified model we implement the desorption barrier as follows. 
A particle diffuses in the bulk with
a rate $w_{\alpha}=\Gamma_{\alpha} \exp \left ( - E_{\alpha} / kT \right )$,
and it jumps out of the system at the boundary site with a rate $X_{\alpha} =
M_{\alpha} \exp \left ( -V_{\alpha} /kT \right ) $, where $V_{\alpha} >
E_{\alpha}$. As before, the pre-factor $M_{\alpha}$ is chosen such that the
desorption rate changes over a substantial range in the experimental
temperature range. This is ensured by choosing $M_{\alpha}=\exp
(V_{\alpha}/kT_f)  $, where $T_f$ is the final temperature. The other rules
remain same as before.

\begin{figure}
\includegraphics[scale=1.0,angle=0]{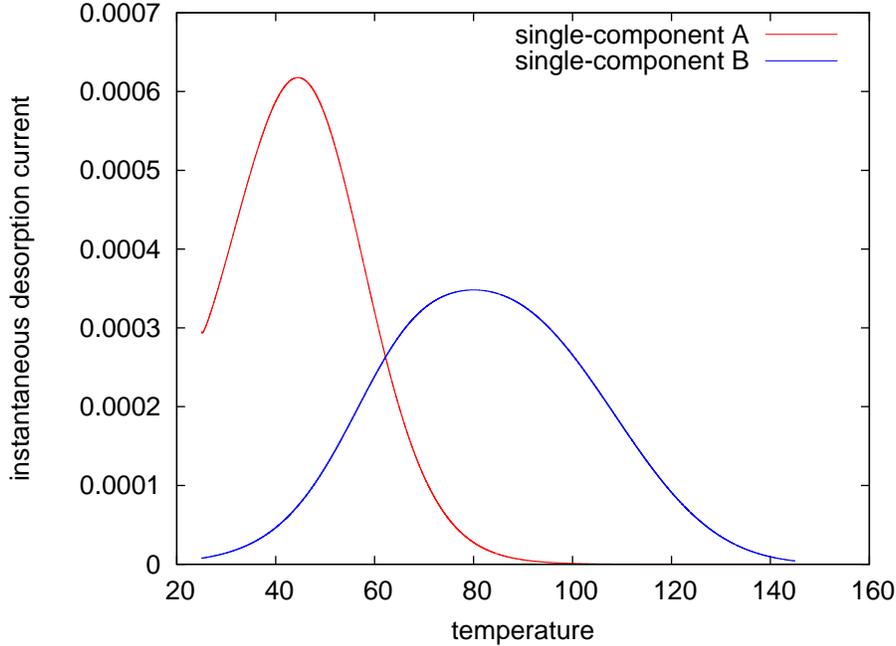}
\caption{Instantaneous desorption current as a function of temperature (in  $^\circ$C)
for single-component diffusion with desorption barrier. We have used $L=1000$, 
$\lambda=5 \times 10^{-5}$ $^{\circ}$C per unit time,
$E_A=1.6$ kJ/mol, $V_A=68.1$ kJ/mol, $E_B=33.2$ kJ/mol, $V_B=99.8$ kJ/mol. }
\label{fig:exitsingle}
\end{figure}

In presence of a single component, with two different rates for bulk diffusion
and boundary desorption,  the 
time-evolution of the local density $\rho_x (t)$ at the
bulk is governed by the equation 
\begin{equation}
\frac{\partial \rho_x (t)}{\partial t} = w_{\alpha} (t) \left ( \rho_{x+1} (t) +
\rho_{x-1} (t) - 2 \rho_{x}(t)  \right ) \quad\quad x \neq 1,L.
\label{eq:exitbulk}
\end{equation} 
At the boundary one has
\begin{eqnarray}
\nonumber
\frac{\partial \rho_1(t)}{\partial t} &=& w_{\alpha} (t) \left ( \rho_2 (t)
-\rho_1 (t) \right ) - X_{\alpha}(t) \rho_1 (t) \\
\frac{\partial \rho_L(t)}{\partial t} &=& w_{\alpha} (t) \left ( \rho_{L-1} (t)
-\rho_L (t) \right ) - X_{\alpha}(t) \rho_L (t) .
\label{eq:exit}
\end{eqnarray}
Note that in the above equation, the boundary densities
couple with the new desorption rate $X_{\alpha}$. In Eqs. \ref{eq:exitbulk}
and \ref{eq:exit} it is not possible to remove
 the explicit time-dependence of both $w_{\alpha}$ and  $X_{\alpha}$ by
performing a single scaling transformation on the time-variable, 
as done in the previous section. Therefore, a solution in closed form
similar to   Eq. \ref{eq:rho} cannot be obtained for this case. In order to 
solve the above set of equations numerically, we write them in the compact
form
\begin{equation}
\tilde {\rho}(t+\delta t) = {\mathcal D}(t) {\tilde\rho} (t)
\label{eq:compact}
\end{equation}
where $\delta t$ is an infinitesimal increment in time and 
the $\tilde{\rho}(t)$ is a column vector, defined as
\begin{equation}
{\tilde \rho}(t) =
\left ( 
\begin{array}{c}
\rho_1(t) \\
\rho_2(t) \\
\rho_3(t) \\
.\\
.\\
.\\
\rho_{L-1}(t) \\
\rho_L(t)
\end{array}
\right ).
\end{equation}
 ${\mathcal D}(t)$ is the ``transfer matrix'', given by
\begin{equation}
{\mathcal D}(t)
=\left ( \begin{array}{ccccccc}
1-X_{\alpha}(t)-w_{\alpha}(t) & w_{\alpha}(t) & 0 &0 &...&0 &0 \\
w_{\alpha}(t) & 1-2w_{\alpha}(t) & w_{\alpha}(t) & 0& ...&0&0 \\
0 & w_{\alpha}(t) & 1-2w_{\alpha}(t) & w_{\alpha}(t) &0& ... &0\\
...& ...&...&...&...&...&...\\
0 & 0&...&...& 0 & w_{\alpha}(t) & 1-X_{\alpha}(t)-w_{\alpha}(t) 
\end{array} \right ) 
\end{equation}
Starting from an initial condition $\tilde{\rho}(0)$
 it is then possible to obtain the density
profile at time $t$ by repeatedly applying  (\ref{eq:compact}). As explained
in the previous section, we choose a homogeneous initial condition.
Using the resulting time-dependent local density
profile $\rho_x(t)$ we compute the instantaneous desorption current
 $J_{\alpha}(t)=w_{\alpha}(t)\left ( \rho_1(t)+\rho_{L-1}(t) \right )$. 
We adjust the parameters $\lambda$, $E_{\alpha}$ and $V_{\alpha}$ such that
the peak of $J_{\alpha}(t)$ is now placed close to 
 the experimentally observed temperature. We present
our plot of $J_{\alpha}(t)$ as a function of temperature in Fig.~\ref{fig:exitsingle}. 

Therefore, including the boundary effect in the form of a desorption barrier,
we can find good quantitative agreement with the experimental desorption
temperature for single-component loading. Note that to compute the desorption
profile in this case, we did not have to do time-consuming Monte Carlo
simulations. The quantitatively correct desorption profile  was obtained by
carrying out an exact numerical integration of (\ref{eq:exitbulk}) and (\ref{eq:exit}).
This computational advantage
enables us to freely explore the various parameter regimes in our model. 
The position of
the peak is sensitive to the choice of these parameters--a large value of the
energy barriers shifts the peak-position to higher temperature, as seen from
the two curves presented in Fig. \ref{fig:exitsingle}. The choice of 
$\lambda$ also affects the desorption profile. A large value of $\lambda$ 
means a fast increase in the diffusivity and as a result the current shows a 
rapid growth as a function of $T$ and the effect of the depletion (see section
$2$) is not felt until higher temperature when the current falls off quickly.
Thus a choice of large $\lambda$ yields a higher desorption temperature. We
verified this in calculations not presented here. On the other hand, 
similar calculations show  that a too small $\lambda$
brings about a qualitative change in the profile by introducing an initial
decay in the current, as discussed in section $2$. Note that $\lambda =0$
corresponds to the equilibrium case where the diffusivity does not change with
time. For such a case one expects an exponential decay of the desorption
current as a function of time. The low-temperature decay of the desorption
current for very small $\lambda$ is nothing but a remnant of this equilibrium
behavior.  
 
\section{Dynamical Monte Carlo simulation for the two-component loading}

In this section we discuss the case where a binary mixture of $A$ and $B$ are
diffusing on the lattice. For this two-component case we perform 
dynamical Monte Carlo simulations 
for  measuring the TPD profile.  We start with a homogeneous density profile, 
which is based on
the assumption that the loading procedure in the experiment would generate
uniform density inside the EUO pore.

Each Monte Carlo time step consists of $(L+1)$
update trials. During each such update trial a lattice bond is chosen uniformly at
random. If the bond lies in the bulk, then the occupancies of the adjacent
sites are updated following the rules described in section $2B$. If the bond
lies at the boundary and the adjacent boundary site is occupied by an
$A$($B$) particle, then the boundary site is emptied with probability
$w_{A}$($w_B$),  with jump rates now understood as dimensionless
jump probabilities for one jump attempt. 
More details on the simulation procedure has been given
in~\cite{sg}.

In our earlier work without desorption barrier
our simulations showed that the presence of $B$ particles indeed raises the desorption
temperature for the $A$ component. The desorption temperature for $A$ in the binary mixture is
substantially higher than for the single-component case, as reported in the
experiment. This shows that the strongly adsorbed $B$ particles can successfully
trap the weakly adsorbed $A$ particles. However, at low temperature, some $A$
particles are still found to escape. We have shown in~\cite{sg} that this
loss can be prevented by starting with an initial condition where no untrapped
$A$ particles are present near the boundary of the lattice.

\begin{figure}
\includegraphics[scale=0.5,angle=0]{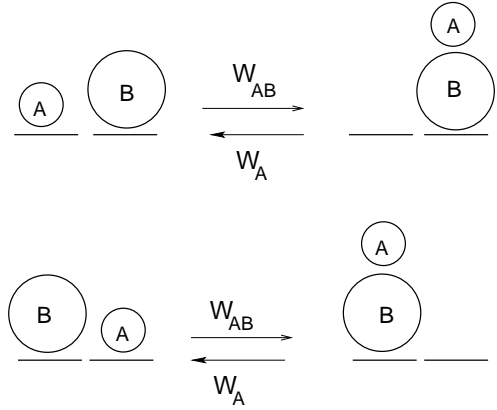}
\caption{$A$ particles can cross over $B$ particles with a small rate
when the single-file condition is relaxed.}
\label{fig:cross}
\end{figure}

\begin{figure}
\includegraphics[scale=1.0,angle=0]{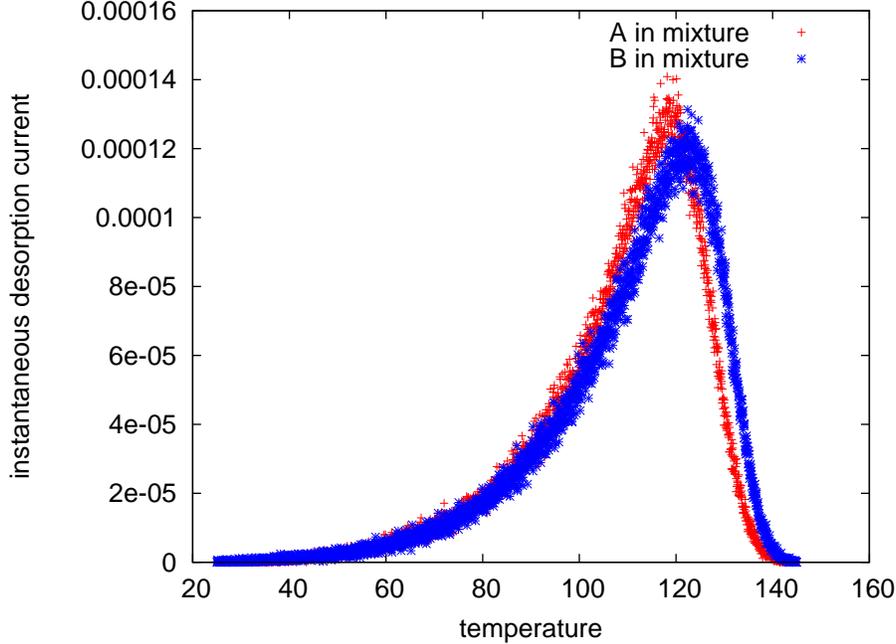}
\caption{ Desorption profile for $A$ and $B$ with a relaxed single-file
condition and 
without desorption barrier. We have used $E_{AB}=183$ kJ/mol and 
$\Gamma_{AB}=exp(E_{AB}/kT_f)$. The other parameters remain same as
 in Fig. \ref{fig:single}.}
\label{fig:wide}
\end{figure}

A closer examination of the experimental data (Fig. \ref{fig:fig5}) reveals
that near the desorption peak the propane current is slightly higher than the
toluene current. However, within a strict single-file condition this 
would not be
possible. If the propane molecules can desorb only after the toluene molecules
have desorbed, it is not possible to obtain a larger propane peak for an 
initial loading with an equimolar mixture. Therefore, to explain the experimental
observation, we had to slightly relax the single-file condition by allowing
`crossing-events' as shown in Fig. \ref{fig:cross} (see~\cite{sg} for details).
In this relaxed single-file environment we retrieve a 
higher $A$-current than the $B$-current near the peak (see Fig.
\ref{fig:wide}).


\begin{figure}
\includegraphics[scale=1.0,angle=0]{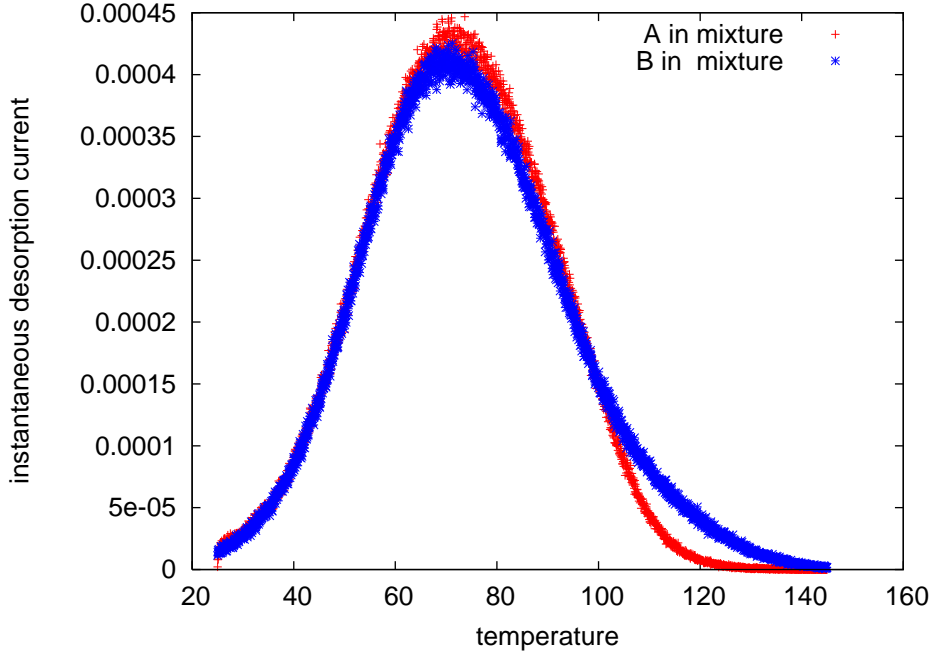}
\caption{Desorption profile of $A$ an $B$ with a relaxed single-file
condition and with desorption barrier.
 We have used $E_{AB}=83.1$ kJ/mol. The other parameters remain
 same as in Fig. \ref{fig:exitsingle}. Note that the inclusion of the
desorption barrier strongly changes the position of the peak and now the peak
is obtained at a temperature which is close to the experimental value.}
\label{fig:exitwide}
\end{figure}

Now let us see what happens when the effect of the desorption barrier is
 included in the model. Since our earlier study reveals that some violation of
the single-file condition takes place inside the channels, we carry out our
simulation in a `relaxed' single-file environment as in~\cite{sg}. The
simulation procedure remains essentially the same but the desorption now takes
place with the modified rate $X_{\alpha}$. We present our data in Fig.
\ref{fig:exitwide}.

 The main observation is that we find a larger propane peak than toluene peak as in the
experiment. Secondly, both peaks occur at $70 ^{\circ}$C which is close to the
experimental value, suggesting that our boundary desorption barrier approach,
along with weak violation of the single-file condition, captures major
physical mechanisms that determine the diffusion and desorption 
of the binary mixture of molecules.

Further comparison between the experimental data in Fig. \ref{fig:fig5} and 
our results in Fig. \ref{fig:exitwide} 
shows that not all the aspects of the experimental data are captured
within our model.  The experiment shows that 
the presence of toluene strongly dominates the desorption process of propane
but their profiles do not become identical. One can see that the toluene desorption
takes place over a wider temperature range than propane. In our model 
find that the tail of the profile for $B$ indeed stretches to higher
temperature as seen in experiment. However, in contrast to the experiment, 
over a substantial range of
temperatures the two profiles lie rather close, unlike Fig. \ref{fig:fig5}.
In order to resolve this remaining discrepancy, one may have to take into account
interactions between the molecules in more detail.
As discussed in section $3$, the interaction between the molecules
gives rise to collective effects like concerted movement of
molecular clusters~\cite{sholl}. It would be interesting to investigate whether 
some such effect is responsible for the relatively smaller width of the propane 
desorption profile observed in experiment.

\section{Conclusions}

We have attempted to describe the diffusion of hydrocarbon
molecules in a narrow zeolite channel using the
two-species symmetric exclusion process. Our earlier studies showed that using
activated diffusion of
hard-core particles on a lattice and assuming an Arrhenius dependence of the
diffusivities on temperature one can explain major qualitative features of
the experiment~\cite{sg}. However, this model was not fit for a quantitative
comparison with the experiment data as the desorption temperature obtained from
this model was much higher than the experimental observation. In this paper,
we have shown that in order to predict a realistic desorption temperature,
one has to take into account boundary effects in the form of a desorption
barrier: Apart from an activated diffusion in the bulk of the lattice, the
particles must overcome an additional energy barrier to desorb from the
boundary. In a real system such a barrier comes from the attractive van der
Waals interaction between the molecules and the pore wall.

The large energy barrier at the boundary has been found to have a strong
influence for shorter zeolite channels~\cite{vasenkov}.
 A molecular dynamics study of  
the tracer exchange of guest molecules between a zeolite
crystal and the surrounding gas-phase shows that if the channel is
short, then the desorption barrier gives rise to a flat density profile
inside the channel. The appearance of a flat profile indicates
that the equilibrium within the crystal is approached fast compared to the
time needed to reach the concentration necessary to maintain equilibrium with
the surrounding gas phase. Because of the large energy barrier at the
pore exit, the system needs a long time to establish equilibrium with the
surrounding gas-phase. In~\cite{vasenkov} it was argued that when calculating
the diffusion coefficients for short zeolite pores one should take into
account the presence of such flat intra-crystalline density profiles. This
could help to overcome the observed 
discrepancies with the diffusion coefficients obtained by microscopic methods.  
Our study shows that the desorption barrier is important also for long zeolite
channels under non-equilibrium conditions since it changes significantly the
desorption temperature in the TPD measurement.

From a modeling perspective we point out that
the quantitative agreement between our Monte Carlo simulation results for the
two-component case and the experimental data demonstrates that even
without considering the details of the interactions present at the molecular
level (as in molecular dynamics simulation) it is possible to explain the
experimental results of~\cite{snurr}
 quantitatively with a good degree of accuracy.  This suggests that adding
simple coarse-grained interaction potentials would allow for a very detailed
quantitative description of the experimental process.
 This is an important conclusion because a molecular
dynamics simulation method is computationally much more demanding than a 
Monte Carlo method and steady-state conditions are usually out of the range of 
accessibility of molecular dynamics simulations.

\section{Acknowledgments}
Financial support by the Deutsche Forschunsgemeinschaft within the priority
program SPP1155 is gratefully acknowledged.

\end{document}